\documentclass{PoS}

\usepackage{bbm}

\title{New developments for dual methods in lattice field theory at non-zero density}

\ShortTitle{New developments for dual methods in lattice field theory at non-zero density}

\author{{Christof Gattringer$\,$}     \thanks{This work has been partly supported by the Austrian Science Fund 
FWF, DK {\it Hadrons in Vacuum, Nuclei, and Stars} (FWF DK W1203-N16),
the Research Executive Agency (REA) of the European Union 
under Grant Agreement number PITN-GA-2009-238353 (ITN STRONGnet), by ERC {\it Hadron Physics 2}, 
by DFG TR55, ``{\sl Hadron Properties from Lattice QCD}'' 
and by the Austrian Science Fund FWF, Grant.\ Nr.\ I 1452-N27.}\\
        Karl-Franzens-Universit\"at, 8010 Graz, Austria\\
        E-mail: \email{  christof.gattringer@uni-graz.at}}

\abstract{In recent years the complex action problem of lattice field theory at 
finite density was overcome for several system by mapping them to dual variables
(flux lines and surfaces). We illustrate this mapping for the case of the U(1) 
gauge Higgs system and present some representative physics results for this 
model. Conceptual challenges such as spectroscopy in the dual approach, as 
well as algorithmic developments are discussed and related ideas for systems with 
fermions are addressed. Models where the dual 
variables approach solves the complex action problem can serve as reference 
systems for other approaches to finite density lattice field theory and we discuss 
some examples. Finally we address the strategy of a partial
dualization in certain limits, e.g., for strong coupling and large mass.}

\FullConference{31st International Symposium on Lattice Field Theory - LATTICE 2013\\
		July 29 - August 3, 2013\\
		Mainz, Germany}

\begin{document}

\section{Introductory remarks}
Numerical simulations of lattice field theories at finite density\footnote{Up to a few 
exceptions such as QCD with gauge group SU(2).} have to deal with the ''complex action problem'': 
For finite chemical potential $\mu$ the action $S$ becomes complex and the Boltzmann factor $e^{-S}$ 
can no longer be used as a weight factor in a Monte Carlo simulation. 
The complex action problem appears for both, bosonic and fermionic systems -- 
for the latter the fermion determinant is complex for non-zero $\mu$. 
The appearance of a complex action at finite $\mu$ is not a lattice artifact but a 
genuine feature of many quantum field theories at finite chemical potential.

Many ideas to overcome the complex action problem have been discussed over the years 
and were covered in reviews at the annual lattice conferences and other occasions
\cite{rev1}--\cite{rev10}: Reweighting techniques, series expansions, analytic 
continuation, stochastic methods, density of state techniques,
canonical simulations and other approaches were tested and helped 
to understand aspects of QCD and other lattice field theories at finite density. On the other hand it is also 
well known that many of these approaches have clear limitations for the values of the 
chemical potential they can reach and thus can only be considered in a limited 
range of parameters\footnote{New developments have been presented for stochastic 
approaches \cite{rev1}, \cite{langevin1}--\cite{lefschetz5} and it will be interesting to see 

how far they reach when realistic models and system sizes are considered.}.

A completely different approach is the idea of trying to find a parametrization of the partition sum in
terms of new variables, such that in the transformed partition sum only real and positive contributions
appear also at finite chemical potential $\mu$. Having a reformulation with only real and positive contributions then 
implies that suitably adapted Monte Carlo algorithms can be used to simulate the system at arbitrary $\mu$
and the complex action problem is solved. 

Several different strategies were followed for finding suitable parametrizations, and here we will often refer
to the new representation used as ''dual representation'' or ''dual variables'', although often it is not a 
conventional duality transformation that has been used for rewriting the system. In recent years dual 
representations were found for a wide range of models, to a large extent bosonic models with abelian symmetry 
\cite{bosonfirst}--\cite{bosonlast}, but also some non-abelian cases were studied, and with 
related methods also some fermionic systems were successfully treated \cite{fermionfirst}--\cite{fermionlast}.

For many forms of dual representations matter fields assume the form of oriented loops on the lattice and 
gauge fields turn into surfaces made of plaquettes that can either be closed surfaces or bounded by matter 
flux\footnote{In the next  section we show this explicitly for the U(1) gauge-Higgs system.}. These dual 
variables are subject to constraints, i.e., for the matter loops the flux is conserved at each site of the lattice, 
and for the gauge fields a similar constraint on the links ensures that the gauge surfaces are closed or
bounded by matter flux. Thus for a successful Monte Carlo simulation one needs to find algorithms that 
take into account these constraints and generate only configurations that are admissible, i.e., obey all the 
constraints. For systems with only fluxes, the Prokof'ev -- Svistunov worm algorithm \cite{worm}
and generalizations of it solve this issue. For systems with gauge surfaces local dual updates were
used  initially \cite{greensite,panero,azcoiti}, but more recently generalizations of the worm idea to surfaces were 
implemented successfully \cite{endres1}--\cite{swa2}.

An important aspect of many dual representations is the fact, that not all observables can be rewritten in a 
trivial way to the dual representation. Straightforward are bulk observables which are obtained as derivatives
of the free energy with respect to the parameters of the theory. These derivatives can be evaluated also in the dual
representation and the bulk observables assume the form of moments of suitable averages of dual variables. 
More complicated are $n$-point functions, which are, e.g., needed for spectroscopy calculations. For these
one usually introduces source terms, which, however, change the form of the dual representation: The 
source terms introduce local contributions to the matter loops, i.e., the constraints for the matter flux are violated at the
points where the sources are activated. The resulting representation can be viewed as a more general dual
form with defects at some of the sites. Suitable generalizations of the worm algorithm were discussed  
\cite{wolff2point} and the feasibility of dual spectroscopy at finite $\mu$ was demonstrated
\cite{dualspectroscopy1,dualspectroscopy2}.

Let us conclude these introductory remarks with pointing out possible applications of models where the complex 
action problem  has been solved completely with dual techniques. A first application is of course the study of interesting
non-perturbative phenomena in finite density field theories. An important example is the so-called Silver Blaze
phenomenon \cite{cohen}, which is the fact that as long as the chemical potential $\mu$ is smaller than the mass $m_0$
of the lightest excitation, physics is independent of $\mu$, and only at $\mu = m_0$ condensation sets in. The Silver Blaze 
phenomenon was studied in various theories recently 
\cite{langevin14,langevin15,lefschetz3,silverblaze,bosonlast,owe1,owe2,wipf1,wipf2}. Besides ab-initio studies of density related 
phenomena in quantum field theories, the systems that can be solved with dual methods may also serve as reference
models to test other approaches to finite density on the lattice. Examples where stochastic methods were compared to
dual results can be found in \cite{rev1,langevin8,lefschetz1,lefschetz3}, and tests of series expansions in $\mu$ were
presented in \cite{seriestest1,seriestest2}.

\section{Dual representation for the U(1) gauge Higgs system}

In this section we discuss the mapping of the U(1) gauge Higgs model to a dual representation. 
This example is characteristic for similar mappings in other theories with matter and gauge fields 
and illustrates the essential steps, as well as the typical structure of the dual representation. 
 
The degrees of freedom are gauge fields $U_{x,\nu} \in$ U(1) on the links of the lattice
and  a charged scalar Higgs field $\phi_x\ \in\ \mathbbm{C}$, attached to the sites.  
We consider a 4-dimensional lattice with volume $V_4 = N_s^3 \times N_t$ and periodic 
boundary conditions for both fields. The total action is a sum of gauge and matter parts,
$S = S_G + S_M$. For the gauge action we  use the standard Wilson form,
\begin{equation}
S_G\ =\ -\frac{\beta}{2} \sum_x \sum_{\nu < \rho} 
\left[ U_{x,\nu\rho}\ +\ U^*_{x,\nu\rho} \right]\ ,
\label{gauge_action}
\end{equation}
where $U_{x,\nu\rho} = U_{x,\nu}U_{x+\hat{\nu},\rho}U^*_{x+\hat{\rho},\nu}U^*_{x,\rho}$ and 
$\beta$ is the inverse gauge coupling. The action for the matter fields is given by
\begin{equation}
S_M = \sum_x \left[ M^2 |\phi_x|^2\ +\ \lambda |\phi_x|^4\ -\
\sum_\nu \left( e^{-\mu \delta_{\nu 4}} \, \phi_x^*U_{x,\nu}\phi_{x+\hat{\nu}} \, + \, e^{\, \mu \delta_{\nu 4}} \,
\phi_x U^*_{x,\nu}\phi_{x+\hat{\nu}}^* \right) \right]\ .
\label{action_higgs}
\end{equation}
The parameter $M^2$ denotes $8+m^2$, where $m$ is the bare mass parameter and
$\lambda$ is the quartic coupling. A non-zero chemical potential
$\mu$ gives a different weight to forward and backward temporal hops (the temporal direction is the 4-direction and
the Kronecker delta $\delta_{\nu 4}$ switches on the chemical potential terms only for $\nu = 4$) and
it is obvious that for $\mu \ne 0$ the action becomes complex. In the standard representation the gauge Higgs model
thus has a complex action problem. 

The partition sum $Z = \int D[U] D[\phi] e^{-S_G - S_M}$ is 
given as an integral over all field configurations.  For the Higgs field the measure is a 
product over all lattice points $x$, and we use polar coordinates  $\phi_x = r_x e^{i \theta_x} 
$ for integrating each $\phi_x$ in the complex plane. The U(1) gauge variables 
$U_{x,\nu} = e^{i\varphi_{x,\nu}}$ at each link are integrated over the unit circle such that the path integral reads
\begin{equation}
Z  = \int\!\! D[U] D[\phi] \, e^{-S_G - S_M} = \left( \prod_{x,\nu}\! \int_{-\pi}^\pi \!\!\frac{d \varphi_{x,\nu}}{2 \pi} \!\!\right) \!\!
\left(\! \prod_{x}\! \int_{-\pi}^\pi\!\! \frac{d \theta_{x}}{2 \pi}\!\! \int_0^\infty\!\!\! dr_x \, r_x \!\!\right) \! e^{-S_G - S_H}.
\label{Zconvent}
\end{equation}
 
We begin the mapping to the dual representation with the Higgs part of the problem
and define the partition sum of the Higgs system in a gauge background as
\begin{equation}
Z_H \; = \; \int\!\! D[\phi] e^{-S_M} = \int\!\! D[\phi]\! \left( \prod_{x,\nu} e^{ \, e^{-\mu \delta_{\nu 4}} \, 
\phi_x^\star U_{x,\nu} \phi_{x+\widehat{\nu}}} 
e^{ \, e^{\, \mu \delta_{\nu 4}} \, \phi_x U_{x,\nu}^\star \phi_{x+\widehat{\nu}}^\star} \! \right) \!\! \left(\prod_x B(|\phi_x|^2)\!\right) ,
\end{equation}
where we have reorganized the nearest neighbor terms and $B(r^{\, 2}) \equiv \exp(-M^2 r^2 - \lambda r^4)$.
The Boltzmann factors for the nearest neighbor terms (use $U_{x,\nu}^{\;\;\star} = U_{x,\nu}^{\;\;-1}$) are now expanded,
\begin{eqnarray}
\hspace*{-2mm}&& \prod_{x,\nu} \exp\!\left( e^{-\mu \delta_{\nu 4}} \, \phi_x^\star U_{x,\nu}    \phi_{x+\widehat{\nu}}\right) 
 \exp\!\left(  e^{\, \mu \delta_{\nu 4}} \, \phi_x U_{x,\nu}^\star \phi_{x+\widehat{\nu}}^\star \right)  =   
\nonumber \\
\hspace*{-2mm}&& \left( \prod_{x,\nu} \sum_{n_{x,\nu}=0}^\infty  \; \sum_{\overline{n}_{x,\nu}=0}^\infty \right) \!\!
\left( \prod_{x,\nu}\!  \frac{ \, e^{-\mu n_{x,4} \delta_{\nu 4}} \; e^{\,\mu \overline{n}_{x,4} \delta_{\nu 4}} \; \; U_{x,\nu}^{\;\;n_{x,\nu}} \, 
{U_{x,\nu}^\star}^{\overline{n}_{x,\nu}}}{n_{x,\nu}! \, \overline{n}_{x,\nu}!} \right) \!
\left( \prod_{x,\nu}\! \Big(\phi_x^\star \phi_{x+\widehat{\nu}}\Big)^{n_{x,\nu}} \, 
\Big(\phi_x \phi_{x+\widehat{\nu}}^\star\Big)^{\overline{n}_{x,\nu}} \! \right) \!= 
\nonumber \\
\hspace*{-6mm}&& 
\sum_{\{ n, \overline{n}\}} \!\!
\left( \prod_{x,\nu}\! \frac{U_{x,\nu}^{\;\;n_{x,\nu}-\overline{n}_{x,\nu}}}{n_{x,\nu}! \, \overline{n}_{x,\nu}!} \right)
\left( \prod_{x} e^{-\mu \, ( n_{x,4} \, - \, \overline{n}_{x,4} )}  \right) 
\left(\!  {\phi_x^{\star}}^{\; \sum_\nu [ n_{x,\nu} + \overline{n}_{x-\widehat{\nu},\nu} ] } 
\, {\phi_x}^{\sum_\nu 
[ \overline{n}_{x,\nu} + n_{x-\widehat{\nu},\nu} ] }  \right) ,
\end{eqnarray}
where the expansion variables $n_{x,\nu}$ and $\overline{n}_{x,\nu}$ are non-negative integers attached to the links of the lattice. 
By $\sum_{\{n,\overline{n}\}}$ we denote the sum over all configurations of the expansion variables $n_{x,\nu}, 
\overline{n}_{x,\nu} \in [0,\infty)$. The partition sum of the Higgs field now reads
\begin{eqnarray}
Z_H & = & \sum_{\{ n, \overline{n}\}}  \!\!
\left( \prod_{x,\nu}\! \frac{U_{x,\nu}^{\;\;n_{x,\nu}-\overline{n}_{x,\nu}}}{n_{x,\nu}! \, \overline{n}_{x,\nu}!} \right) 
\left( \prod_{x} e^{-\mu \, ( n_{x,4} \, - \, \overline{n}_{x,4} )}  \right) 
\left(\! \prod_x \int_{-\pi}^\pi \frac{d\theta_x}{2\pi} e^{-i\theta_x \sum_\nu [ n_{x,\nu}  - \overline{n}_{x,\nu} - 
( n_{x-\widehat{\nu},\nu}   - \overline{n}_{x-\widehat{\nu},\nu}) ] }\! \right)  \nonumber \\
& &\hspace{10mm} \times
\left(\prod_x \int_{0}^\infty\!\!\! dr_x \; r_x^{1 + \sum_\nu [ n_{x,\nu} + n_{x-\widehat{\nu},\nu} 
+ \overline{n}_{x,\nu} + \overline{n}_{x-\widehat{\nu},\nu} ] }  \; B\big(r_x^{\,2} \big) \right)\! . 
\end{eqnarray}
The integrals over the phase give rise to Kronecker deltas, which for notational convenience here
we write as $\delta(n)$. For the integrals over the modulus we introduce the abbreviation
$ P(n) = \int_0^\infty dr \, r^{\,n+1} \,  B\big(r^{\,2})$.
The $P(n)$ can be computed numerically and are pre-stored for determining the
acceptance in the Metropolis steps of the Monte Carlo. The partition sum for the Higgs field is
\begin{eqnarray}
Z_H & = & \sum_{\{ n, \overline{n}\}}\!\!  
\left( \prod_{x,\nu}\! \frac{U_{x,\nu}^{\;\;n_{x,\nu}-\overline{n}_{x,\nu}}}{n_{x,\nu}! \, \overline{n}_{x,\nu}!} \right) 
\left( \prod_{x} e^{-\mu \, ( n_{x,4} \, - \, \overline{n}_{x,4} )}  \right) 
\left(\! \prod_x \delta\!\left( \sum_\nu \big[ n_{x,\nu}  - \overline{n}_{x,\nu} - 
( n_{x-\widehat{\nu},\nu}   - \overline{n}_{x-\widehat{\nu},\nu}) \big] \! \right)\!\right)
\nonumber \\
& \!\!  \!\!& \hspace{10mm} \times
\left(\! \prod_x  P\!\left( \sum_\nu \big[ n_{x,\nu} 
+ \overline{n}_{x,\nu} + n_{x-\widehat{\nu},\nu} +  \overline{n}_{x-\widehat{\nu},\nu} \big]  \right) \right)\! . 
\label{intermediate}
\end{eqnarray}
In this form the Higgs fields are integrated out and the partition sum is a sum over  
configurations of the $n$ and $\overline{n}$. The allowed configurations 
of the $n$ and $\overline{n}$ are subject to local constraints at each 
site $x$ enforced by the Kronecker deltas, i.e., at each site $x$ the variables
 must obey  $\sum_\nu [ n_{x,\nu}  - \overline{n}_{x,\nu} - 
( n_{x-\widehat{\nu},\nu}   - \overline{n}_{x-\widehat{\nu},\nu}) ]  = 0$. We stress that in the 
new form (\ref{intermediate}) the chemical potential enters only via the real factors 
$e^{-\mu \, ( n_{x,4} \, - \, \overline{n}_{x,4} )}$ which is the key step for solving the complex
action problem.

The structure of the constraints can be simplified by introducing new variables
$l_{x,\nu} \in (-\infty,\infty)$ and $\overline{l}_{x,\nu} \in [0,\infty)$,  which are related to the old variables by
$n_{x,\nu} - \overline{n}_{x,\nu} = l_{x,\nu}$ and
$n_{x,\nu} + \overline{n}_{x,\nu} = |l_{x,\nu}| + 2 \overline{l}_{x,\nu}$,
and the sum over all configurations of the $n,\overline{n}$ variables 
can be replaced by a sum over $l$- and $\overline{l}$-configurations.
The partition sum turns into
\begin{eqnarray}
Z_H & = & \sum_{\{ l, \overline{l}\}}  \left( \prod_{x,\nu}\! \frac{U_{x,\nu}^{\;\;l_{x,\nu}}  }{(|l_{x,\nu}| + \overline{l}_{x,\nu})! \, \overline{l}_{x,\nu}!} \right) 
\left( \prod_{x} e^{-\mu \, l_{x,4} }  \right)
\left(\! \prod_x \delta\left( \sum_\nu \big[ l_{x,\nu}  -  l_{x-\widehat{\nu},\nu}  \big] \! \right) \right)
\nonumber \\
& \!\!  \!\!& \hspace{10mm} \times
\left(\!\prod_x
P\!\!\left( \sum_\nu \big[ |l_{x,\nu}| +  |l_{x-\widehat{\nu},\nu}| + 2( \overline{l}_{x,\nu} + \overline{l}_{x-\widehat{\nu},\nu}) \big]  \right) \right)\! . 
\label{ZHfinal}
\end{eqnarray}
In the final form (\ref{ZHfinal}) of the Higgs field partition sum, 
which we now refer to as dual representation, the constraints no longer mix the two types of flux variables. 
Obviously only the $l$-fluxes are subject to conservation of flux at each site $x$, i.e., they must obey 
$\sum_\nu [ l_{x,\nu} - l_{x-\widehat{\nu},\nu}] = 0$ for all $x$.

Having mapped the Higgs field partition sum to the flux form  (\ref{ZHfinal}) we now apply similar steps to the gauge fields to obtain the dual 
representation of the full partition sum (\ref{Zconvent}). We write the full partition sum as $Z = \int D[U] e^{-S_G} Z_H$ and find
\begin{eqnarray}
Z &=& \sum_{\{ l, \overline{l}\}}  \left( \prod_{x,\nu}\! \frac{ 1  }{(|l_{x,\nu}| + \overline{l}_{x,\nu})! \, \overline{l}_{x,\nu}!} \right) 
\left( \prod_{x} e^{-\mu \, l_{x,4} }  \right)
\left(\! \prod_x \delta\left( \sum_\nu \big[ l_{x,\nu}  -  l_{x-\widehat{\nu},\nu}  \big] \! \right)\! \right) 
\nonumber
\\
&& \hspace{-3mm} \times
\left(\! \prod_x P\!\left( \sum_\nu \big[ |l_{x,\nu}| +  |l_{x-\widehat{\nu},\nu}| + 2( \overline{l}_{x,\nu} + \overline{l}_{x-\widehat{\nu},\nu}) \big]\!  \right)\! \right)
Z_G[l] \,,
\label{Zfull1}
\end{eqnarray}
where we have interchanged the sum over the flux configurations and the integral over the gauge fields.
The gauge field partition sum with link insertions according to a flux configuration $l$ is defined as
$Z_G[l] = \int D[U] e^{-S_G} \, \prod_{x,\nu}   U_{x,\nu}^{\;\,l_{x,\nu}}$.
The gauge action $S_G$ as defined in (\ref{gauge_action}) is a sum over plaquettes. 
We thus may write the Boltzmann factor $e^{-S_G}$ as a 
product over plaquettes and, as done for the Higgs field, we expand the corresponding exponentials into power series:
\begin{eqnarray}
\prod_{x,\sigma < \tau}\, e^{\frac{\beta}{2} U_{x,\sigma\tau}} \, e^{\frac{\beta}{2} U_{x,\sigma\tau}^\star } &\!\!\!=\!\!\!& 
\sum_{\{ m, \overline{m}\}} \! \left(
\prod_{x,\sigma < \tau }\, \frac{(\frac{\beta}{2})^{m_{x,\sigma\tau} + \overline{m}_{x,\sigma\tau}}}{m_{x,\sigma\tau}! \, \overline{m}_{x,\sigma\tau}!}
 {U_{x,\sigma\tau}}^{m_{x,\sigma\tau}} \; {U_{x,\sigma\tau}^\star}^{\overline{m}_{x,\sigma\tau}} \right)
\label{gaugeboltz}  \\
&\!\!\!=\!\!\!& 
\sum_{\{ m, \overline{m}\}}  \!
\left( \prod_{x,\sigma < \tau }\! \frac{(\frac{\beta}{2})^{m_{x,\sigma\tau} + 
\overline{m}_{x,\sigma\tau}}}{m_{x,\sigma\tau}! \, \overline{m}_{x,\sigma\tau}!} \, \right) \!\!
  \left( \prod_{x,\nu} {\,U_{x,\nu}}^{\sum_{\nu < \alpha} \big[ p_{x,\nu\alpha} - p_{x-\widehat{\alpha},\nu\alpha}\big]
- \sum_{\alpha<\nu} \big[ p_{x,\alpha\nu} - p_{x-\widehat{\alpha},\alpha\nu}\big] }\! \right)\! .
\nonumber
\end{eqnarray}
We introduced the expansion variables $m_{x,\sigma\tau}, \overline{m}_{x,\sigma\tau} \in
[0,\infty)$ attached to the plaquettes, and  by
$\sum_{\{m,\overline{m}\}}$ we denote the sum over all configurations of the  expansion variables.
In the second step we inserted  the explicit expressions for the plaquettes in terms of the link
variables, i.e., $U_{x,\sigma\tau} = U_{x,\sigma} U_{x+\widehat{\sigma}, \tau}
U_{x+\widehat{\tau},\sigma}^\star U_{x,\tau}^\star$, and reorganized the product over powers of
links variables. Here we already introduced  $m_{x,\nu\alpha} - \overline{m}_{x,\nu\alpha}  =
p_{x,\nu\alpha} $. We promote $p_{x,\nu\alpha}  \in
(-\infty,\infty)$ into new dynamical variables, which  together with another set of variables,
$q_{x,\nu\alpha} \in [0,\infty)$, gives the final set of variables we use for the gauge fields. 
The $p$ and $q$ variables are related to the $m$ and  $\overline{m}$ variables via  
$m_{x,\nu\alpha} - \overline{m}_{x,\nu\alpha} = p_{x,\nu\alpha}$  
and $m_{x,\nu\alpha} + \overline{m}_{x,\nu\alpha} = |p_{x,\nu\alpha}| + 2 q_{x,\nu\alpha}$.
We will refer to the variables $p$ as plaquette occupation numbers or simply plaquette variables. 
Using the new variables and inserting the expanded Boltzmann factor (\ref{gaugeboltz}) back into $Z_G$ we find
\begin{eqnarray}
Z_G[l] &\!\! =\!\! & \sum_{\{p,q\}} \left( \prod_{x,\sigma < \tau} \frac{(\frac{\beta}{2})^{|p_{x,\sigma\tau}| + 2 q_{x,\sigma\tau}}}{(
|p_{x,\sigma\tau}| + q_{x,\sigma\tau})! \, q_{x,\sigma\tau}!} \right)
\\
&\!\!\times\!\! & \left(\!\prod_{x,\nu} \int_{-\pi}^\pi \!\!\frac{d \varphi_{x,\nu}}{2\pi} e^{i\varphi_{x,\nu} \left( \sum_{\nu < \alpha} \!\big[ p_{x,\nu\alpha} - p_{x-
\widehat{\alpha},\nu\alpha}\big]
- \sum_{\alpha<\nu}\! \big[ p_{x,\alpha\nu} - p_{x-\widehat{\alpha},\alpha\nu}\big]  + l_{x,\nu} \right) }\!\!\right)\!\! .
\nonumber 
\end{eqnarray}
The integrals in the last product are again representations of Kronecker deltas and give 
rise to constraints that are located at the links of the lattice. The summations over the 
variables $q_{x,\sigma\tau}$ can be done in closed form using the well known series 
representation of the modified Bessel functions
$\sum_{q=0}^\infty \frac{(\frac{\beta}{2})^{|p|+2q}}{(|p|+q)!\, q!} \; = \; I_{|p|}(\beta) \; = \; I_{p}(\beta)$.
Thus we end up with the following representation for $Z_G$:
\begin{equation}
Z_G[l]  \! =\!\sum_{\{p\}} \!\!\left( \prod_{x,\sigma < \tau}\!\! I_{p_{x,\sigma\tau}}(\beta)\!\!\! \right) \!\!\!\!
\left( \prod_{x,\nu} \delta\!\!\left(\sum_{\nu < \alpha} \!\big[ p_{x,\nu\alpha} - p_{x-\widehat{\alpha},\nu\alpha}\big]
- \!\!\sum_{\alpha<\nu} \big[ p_{x,\alpha\nu} - p_{x-\widehat{\alpha},\alpha\nu}\big]  + l_{x,\nu} \!\right)\! \!\right)\!.
\end{equation}
Putting this back into the full partition sum (\ref{Zfull1}) we obtain the final result 
for the dual representation of the partition sum for the U(1) 
gauge Higgs model:
\begin{equation}
Z = \sum_{\{\overline{l},l\}} \sum_{\{p\}} {\cal W}_M[\overline{l},l] \, {\cal W}_G[p] \,
{\cal C}_S[l] \; {\cal C}_L[p,l] \; .
\label{u1dual}
\end{equation}
The weight factors are 
\begin{eqnarray}
{\cal W}_M[\overline{l},l] &\!\!\! = \!\!\!& \prod_{x,\nu} \frac{1}{(|l_{x,\nu}| 
\! + \! \overline{l}_{x,\nu})! 
\overline{l}_{x,\nu}!} 
\, \exp \left( \! - \mu \, \sum_x l_{x,4}  \!  \right) \,
\prod_x P\!\left(\! 
\sum_\nu[|l_{x,\nu}| \!+\! |l_{x-\hat{\nu},\nu}| 
+ 2(\overline{l}_{x,\nu} \!+\! \overline{l}_{x-\hat{\nu},\nu})]\! \right)\!, \nonumber \\
{\cal W}_G[p] &\!\!\!=\!\!\! & \prod_{x,\rho<\nu} I_{p_{x,\rho\nu}}( \beta ) \;.
\label{weight_u1} 
\end{eqnarray}
It is obvious that all weight factors are real and positive and thus the complex action problem
has vanished in the dual representation. 
The $l$ and the $p$ variables are subject to constraints given by
\begin{eqnarray}
{\cal C}_L[p,l] & \!=\! &\prod_x \prod_{\nu=1}^4 \delta \left( \, \sum_{\rho:\nu<\rho}[p_{x,\nu\rho}
- p_{x-\hat{\rho},\nu\rho}] - \sum_{\rho:\nu>\rho}[ p_{x,\rho\nu}
- p_{x-\hat{\rho},\rho\nu} ] + l_{x,\nu}\right) ,
\nonumber \\
{\cal C}_S[l] & \!=\! & \prod_x \delta \left( \, \sum_{\nu=1}^4 [l_{x-\hat{\nu},\nu} -
l_{x,\nu}] \right) . 
\label{constraint_u1} 
\end{eqnarray}
The constraints  come in two parts: We have constraints 
${\cal C}_S[l]$ for the variables $l$ that are based at the sites and enforce the total $l$-flux 
(flux for matter fields) to vanish at each site of the lattice. The second type of constraints
${\cal C}_L[p,l]$ are based on the links and combine $p$ and $l$ variables. They enforce
the total flux from matter and from the plaquettes to vanish at each link. 

\begin{figure}[t]
\centering
\includegraphics[width=105mm,clip]{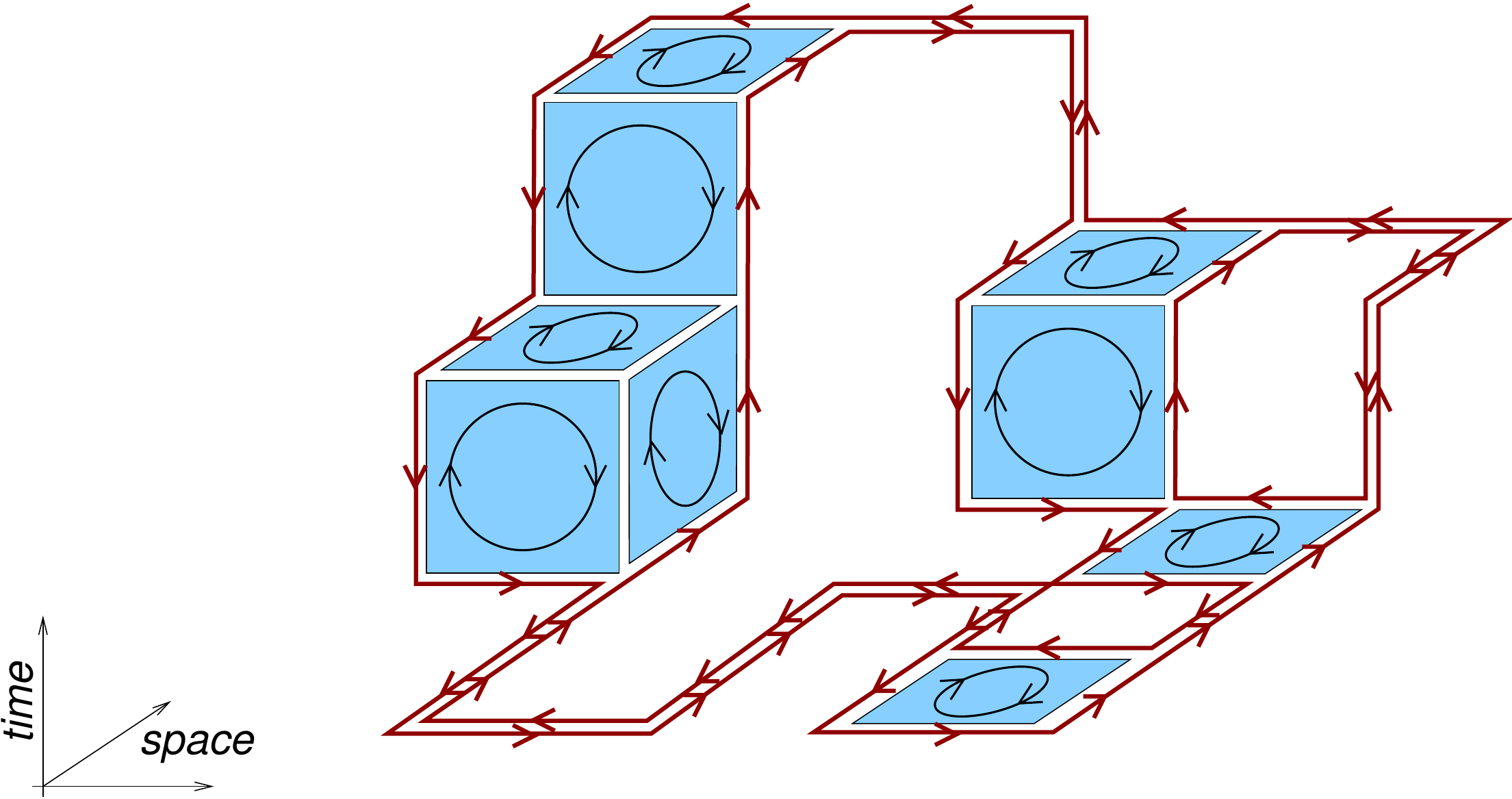}
\caption{Example of an admissible configuration of the dual variables in the U(1) gauge Higgs system.}
\label{configuration}
\end{figure}

We have already stated that typically the dual representation gives rise to loops for the matter degrees of freedom, 
and surfaces for the gauge fields, where the latter are either closed surfaces or surfaces bounded by matter flux. 
This geometrical interpretation is a direct consequence of the constraints and in Fig.~\ref{configuration} we show 
an example of an admissible dual configuration, i.e., a configuration where all constraints are obeyed.  The flux 
variables $l_{x,\nu}$ are represented by full red lines attached to the links with arrows indicating the amount of flux, where the
arrows for negative values point in negative direction (the example shown in  Fig.~\ref{configuration} only has 
$l_{x,\nu} = \pm 1$). Plaquette variables $p_{x,\nu\rho}$ are represented by fluxes drawn onto the corresponding plaquettes.
The example in  Fig.~\ref{configuration} illustrates that at each site the matter flux vanishes, and that at each link the total flux 
from matter and plaquette variables adds up to zero as well.  
 
It interesting to analyze the term  $\exp \left( \! - \mu \, \sum_x l_{x,4}  \!  \right)$  where the chemical potential couples:
In the sum $\sum_x l_{x,4}$ the 4-components of the $l$-flux (which come with positive and negative signs) are added up. Since 
flux is conserved due to the constraints, this term corresponds to $N_t$ times the total net flux of $l$ around the temporal 
direction. Since $N_t$ corresponds to the inverse temperature $\beta$ in lattice units (not to be confused with the inverse gauge coupling here),
we can identify the particle number $N$ via $\exp( \mu \beta N) = \exp \left( \! - \mu \, \sum_x l_{x,4}  \!  \right)$ and
find $N = - \sum_x l_{x,4} \, / \, N_t$.

The fact that the chemical potential couples to the net $l$-flux that winds around the compactified time direction has now to 
be confronted with the link constraints that have to be obeyed. They imply that on each link the $l$-flux is either directly 
matched by one with opposite sign, which, however,
leads to a cancellation of the corresponding factors $e^{\, \pm \mu}$, or compensated by a suitable plaquette value
$p_{x,\nu\rho}$, which, however, only shifts the flux that needs to be compensated 
to a neighboring link. It is easy to see that for obeying the link constraints one is 
either forced to a vanishing particle number $N = - \sum_x l_{x,4} \, / \, N_t = 0$, and thus no $\mu$-dependence, or to attaching infinitely 
many plaquettes to ''shift the problem to infinity'', which gives rise to a configuration with infinite energy. 
Obviously the dual representation knows about the neutrality condition of electrodynamics and for having a non-trivial dependence on 
the chemical potential one needs several flavors with vanishing total charge.
The simplest case is to add a second Higgs field with opposite charge, and in this theory physics at finite chemical potential can be 
studied \cite{swa1,swa2,u1main,bosonlast} (see also below).  

The construction which we discussed here can be generalized to essentially all abelian gauge Higgs systems -- for an example with 
$\mathbbm{Z}_3$ symmetry see \cite{z3}. The question for the existence of a real dual representation for systems with abelian symmetries
was analyzed in a more general setting in \cite{ogilvie1,ogilvie2} and the relation to discrete symmetries was clarified. 
We note that in some special (low dimensional) models exploring discrete symmetries led to an
algorithm (''subset method'') that overcomes the complex action problem already in the conventional representation
\cite{subsets1}--\cite{subsets4}.  

\section{Algorithms}

For a successful Monte Carlo simulation it is mandatory to use an algorithm that produces only 
admissible configurations of the dual variables, i.e., configurations that obey all the constraints. 
The simplest choice is a local algorithm, where the plaquette variables on an embedded 3-cube are 
suitably changed by $\pm 1$, and single plaquettes together with the matter flux at their boundary are changed 
by $\pm 1$ (see, e.g., \cite{greensite,panero,azcoiti,swa1,bosonlast}). However, for many values of the parameters such a 
local algorithm is not very efficient and better strategies are needed.

A powerful tool for the update of dual representations of spin systems or dual versions of lattice 
field theories with only matter fields is the Prokof'ev -- Svistunov worm algorithm \cite{worm}.
In these cases one has only loops of matter flux and the worm algorithm starts with violating 
the constraints at some site and then randomly propagates this defect through the lattice until the worm
returns to the starting site and the violation of the constraint is healed. Each step is accepted 
with a Metropolis decision and the algorithm can be shown to be ergodic. 

It is interesting to note that the idea of the worm algorithm can be generalized to abelian gauge Higgs
systems. Again one violates the constraints at a randomly chosen position on the lattice and 
propagates the defect until the worm closes in a final step where the defect is healed. For systems
with gauge and matter fields two types of constraints appear: Conservation of matter flux at each site
of the lattice and conservation of the combined flux of matter and plaquette variables at each link.
Thus the initial step violates both types of constraints and the individual steps of the generalized 
worm algorithm have to propagate the defects for both constraints simultaneously. 

\begin{figure}[t]
\centering
\includegraphics[width=150mm,clip]{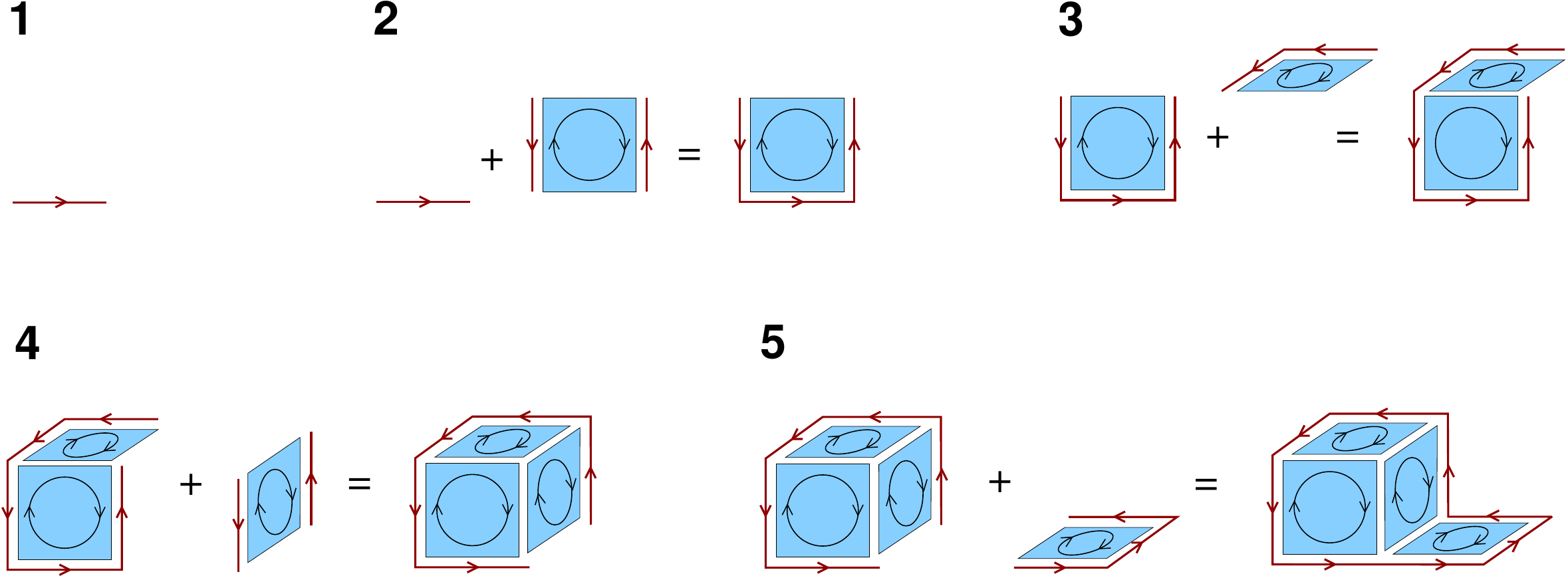}
\caption{Example of updates with the surface worm algorithm (SWA): After randomly placing a link flux (step 1), 
the corresponding defect is propagated through the lattice by attaching plaquettes with two of their sides 
saturated with link flux (steps 2,3,4). The surface worm 
closes by inserting an element that heals the violated constraint (step 5). Each step is accepted with a Metropolis decision.}
\label{SWA}
\end{figure}

A suitably generalization of the worm algorithm to abelian gauge Higgs systems, the so called
''surface worm algorithm'' (SWA), was introduced in \cite{swa1,swa2,u1main}\footnote{For other
attempts to generalize the worm concept to systems with gauge surfaces see 
\cite{endres1,endres2,gaugeworm1,gaugeworm2,gaugeworm3,gaugeworm4}.}. In an initial step
the matter flux at a randomly chosen link is changed by $\pm 1$, which violates the constraint
at the particular link and at the two sites at the two endpoints of the link. Subsequently the defects are 
transported through the lattice with elements of surfaces (''segments'') that consist of plaquettes
where the corresponding $p$-variable and the matter flux at two of the sides of the plaquette is changed 
suitably by $\pm 1$. In a final step the worm randomly adds an element such that all constraints
are healed (see \cite{swa1} for a detailed description). 
The acceptance of each step is governed by a Metropolis decision. In Fig.~\ref{SWA}
we show an example for a short SWA on an initially empty lattice.

\section{Some results for the U(1) gauge Higgs system}

Let us now come to discussing examples of typical results that can be obtained with dual simulations. We use
the U(1) gauge Higgs model with two flavors of opposite charge \cite{u1main,bosonlast}. 
In this section we present results for observables that can be obtained as derivatives of $\ln Z$ with respect to the parameters
of the theory, i.e., masses, couplings and the chemical potential. These derivatives can be taken for both,
the conventional representation, where the corresponding observables can be interpreted, as well as in the dual 
representation to obtain their form for the dual simulation. In the dual world these bulk observables typically
assume the form of weighted sums over the dual variables. Two examples are the plaquette 
$\langle U \rangle$ and the particle number density $\langle n \rangle$:

\begin{equation}
\langle U \rangle  =  \frac{\partial \, \ln\ Z }{6 N_s^3 N_t \, \partial \beta} 
 =    \frac{1}{6 N_s^3 N_t} \left\langle \prod_{x,\rho<\nu} \,  \frac{I_{p_{x,\rho\nu}}( \beta )^\prime}{I_{p_{x,\rho\nu}}( \beta )} \! \right\rangle \; , \;
\langle n \rangle  = \frac{\partial \, \ln\ Z}{6 N_s^3 N_t \, \partial \mu}  =  
\frac{1}{N_s^3 N_t} \left\langle - \sum_x l_{x,4} \! \right\rangle \; ,
\label{observables}
\end{equation}
where the prime in the first equation indicates the derivative with respect to $\beta$ and the 
expressions on the right hand sides are evaluated in the dual representation.

The observables (\ref{observables}) together with $\langle |\phi|^2  \rangle = \partial / \partial M^2 \ln Z  /   N_s^3 N_t$, as well as the 
corresponding susceptibilities $\chi_U$, $\chi_n$ and $\chi_{|\phi|^2}$
were studied in \cite{u1main,bosonlast} for the U(1) gauge Higgs  model with two flavors of opposite 
charge. The mass parameters and the chemical potentials  of the two flavors were set to equal values 
($M^2$ and $\mu$). 

\begin{figure}[t]
\centering
\hspace*{-3mm}
\includegraphics[width=140mm,clip]{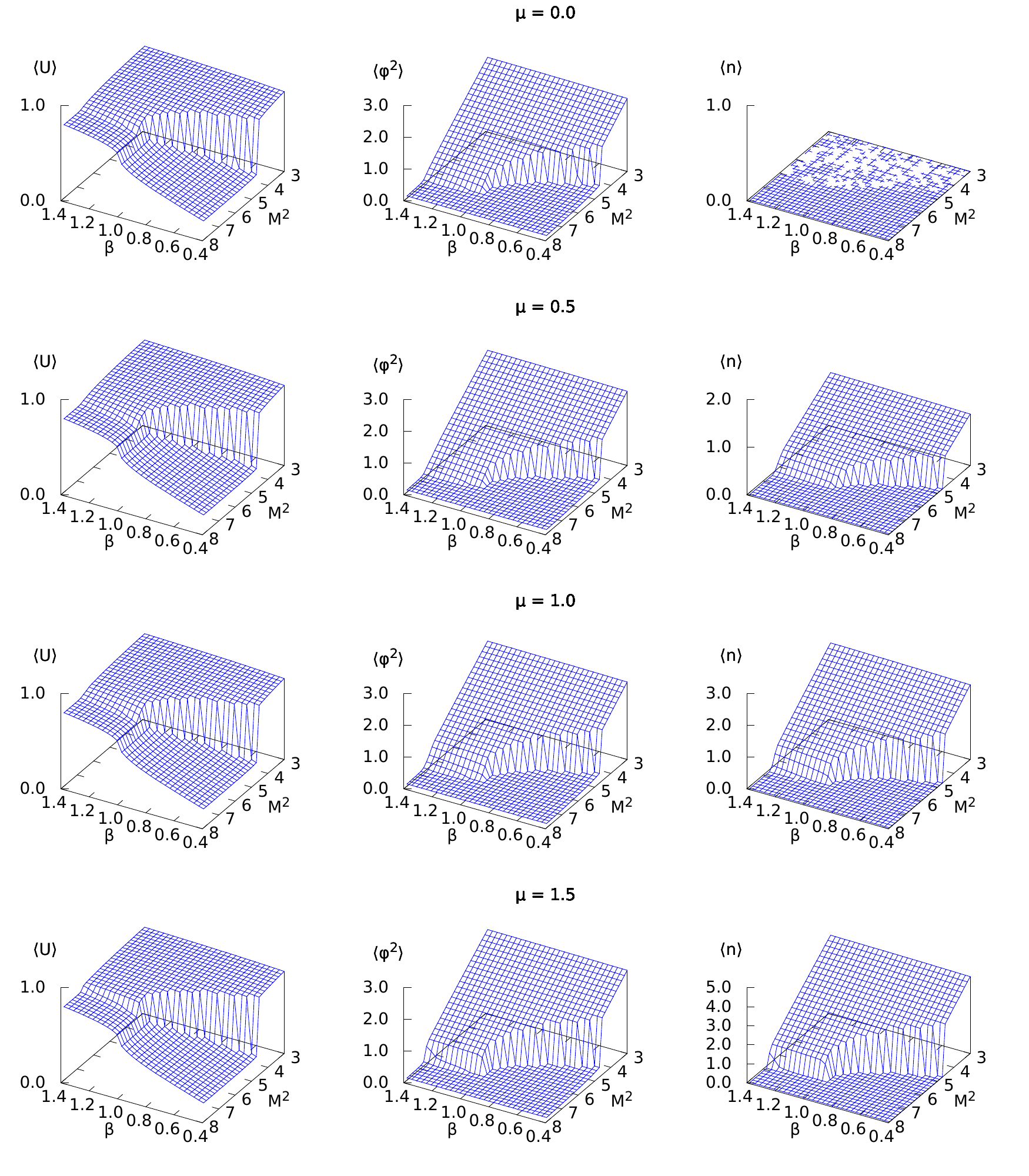}
\caption{The observables $\langle U \rangle$, $\langle |\phi|^2 \rangle$, and 
$\langle n \rangle$ as a function of $\beta$ and $M^2$ for different chemical 
potentials $\mu = 0.0,\,0.5,\,1.0$ and $1.5$. It can be seen how the phase 
boundaries shift with increasing chemical potential.}
\label{muphases}
\end{figure}

For a first scan through parameter space, in Fig.~\ref{muphases} we show the  
observables $\langle U \rangle$, $\langle |\phi|^2 \rangle$, and  $\langle n \rangle$ 
(left to right) as a function of $\beta$ and $M^2$ for different values of the chemical potential,
$\mu = 0.0,\,0.5,\,1.0$ and $1.5$ (top to bottom). With the exception of $\langle n \rangle$ at $\mu = 0$, 
where $\langle n \rangle \equiv 0$, in all observables we see the three phases expected for the
U(1) gauge Higgs model: A confining phase characterized by $\langle U \rangle$ small, 
$\langle  |\phi|^2  \rangle \sim 0$ and $\langle n \rangle \sim 0$, a Coulomb phase 
($\langle U \rangle \sim$ 0.6 -- 0.8,  $\langle  |\phi|^2  \rangle \sim 0$, $\langle n \rangle \sim 0$),
and a Higgs phase ($\langle U \rangle \sim$ 1.0,  $\langle  |\phi|^2  \rangle $ and $\langle n \rangle \sim 0$
growing with $-m^2$). 

The phases are separated by phase boundaries, which partly are signaled
by very abrupt changes of the observables, in particular for the strong first order transition between the 
Higgs- and the confinement phase. Increasing $\mu$ makes the transitions even more pronounced and 
in addition leads to a small shift of the phase boundaries in the $\beta$-$M^2$ plane. 

\begin{figure}[p]
\centering
\hspace*{-3mm}
\includegraphics[width=85mm,clip]{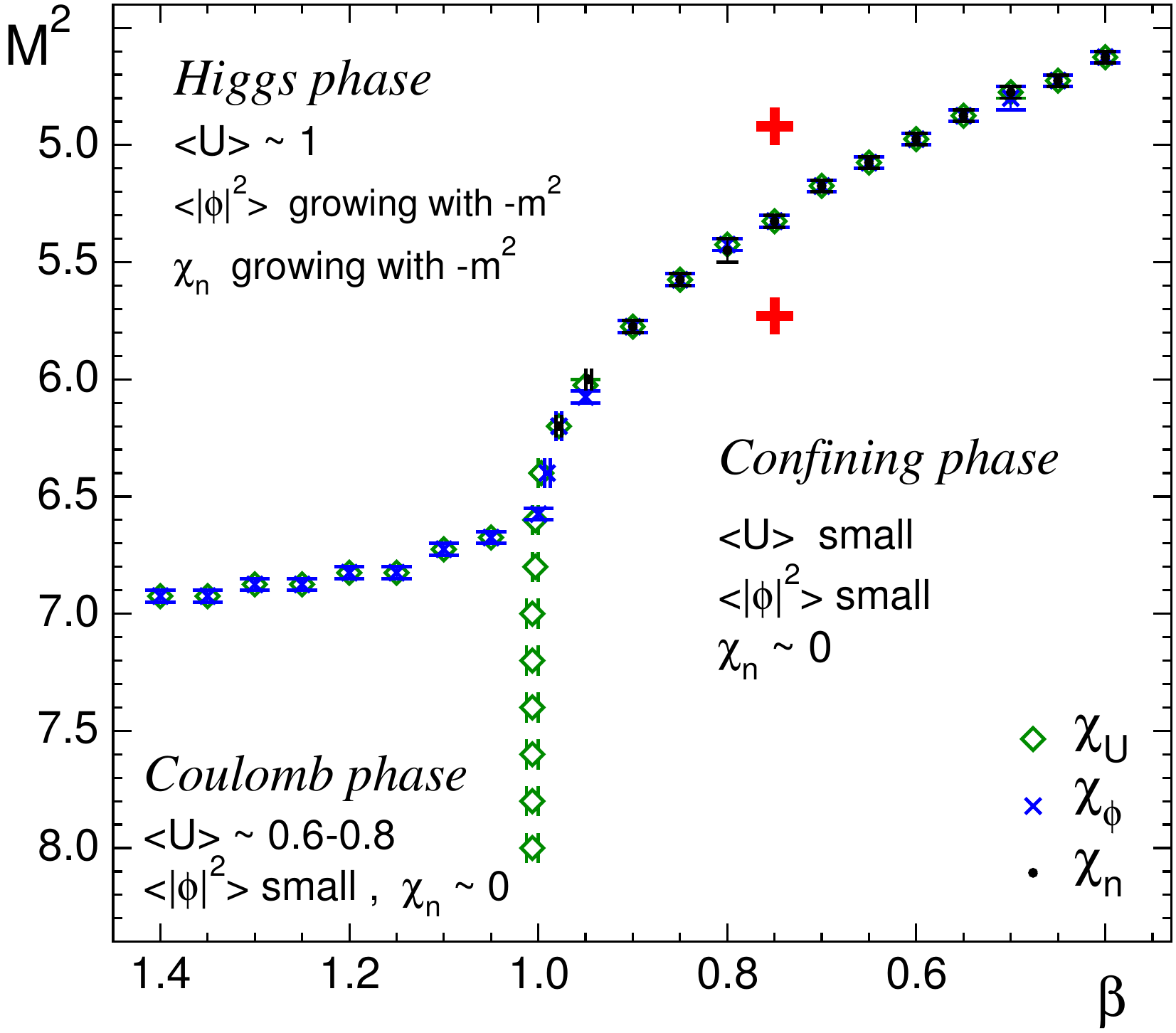}
\caption{Phase diagram in the $\beta$-$M^2$ plane at $\mu = 0$. We show
the phase boundaries determined from the maxima of the susceptibilities $\chi_U$ and $\chi_{\phi}$ and the
inflection points of $\chi_n$.}
\label{phasediagram}
\vskip8mm
\includegraphics[width=150mm,clip]{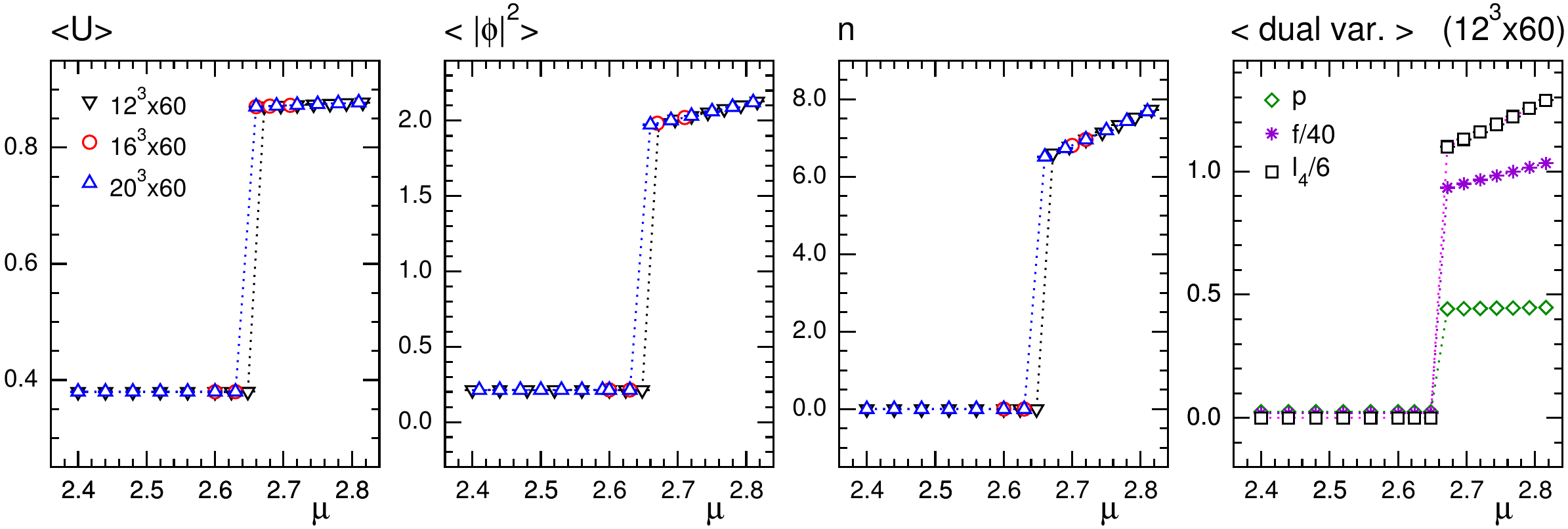}
\includegraphics[width=150mm,clip]{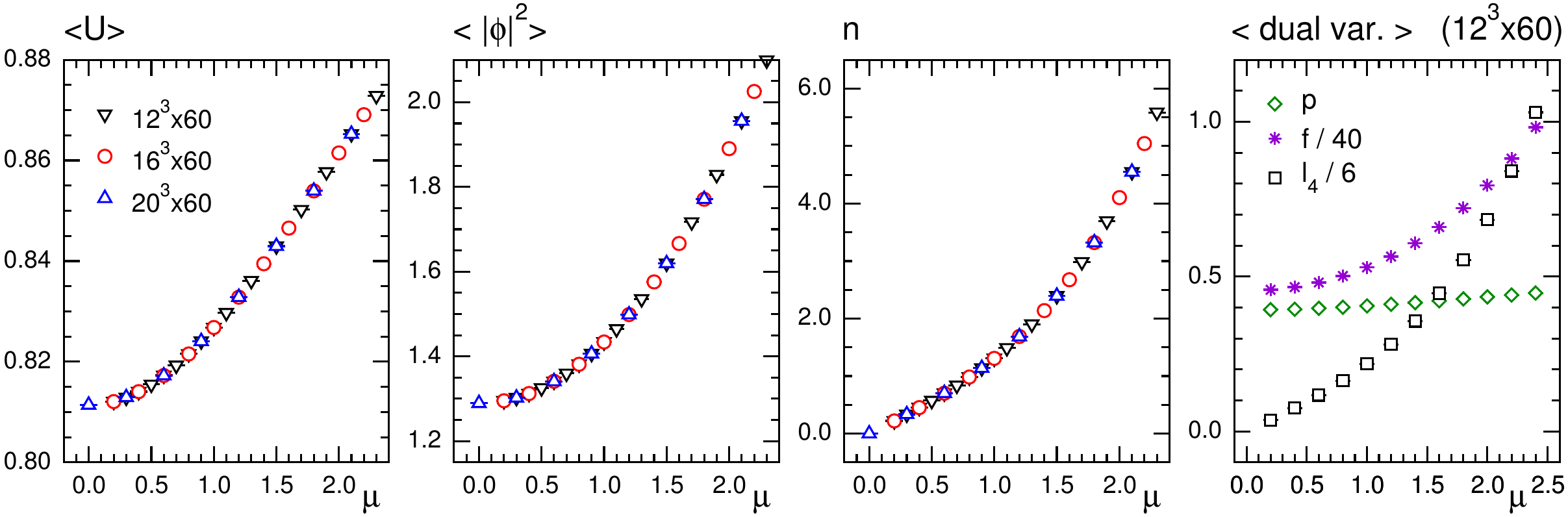}
\caption{The observables $\langle U \rangle$, $\langle |\phi|^2 \rangle$, $n$ 
(left to right) and the occupation numbers of dual variables as a function 
of $\mu$. We compare the results for parameters in the confined phase 
($\beta = 0.75, M^2 = 4.92$, top row of plots) and in the Higgs phase ($\beta = 0.75, M^2 = 5.73$, bottom).}
\label{obs_vs_mu}
\end{figure}

For the case of $\mu = 0$, the phase boundaries were determined in detail \cite{u1main,bosonlast} using the susceptibilities 
$\chi_U$, $\chi_n$ and $\chi_{|\phi|^2}$. The results are shown in Fig.~\ref{phasediagram}. The boundary between the Higgs-
and the confinement phase is  characterized by a strong first order transition, the boundary between Coulomb- and 
confining phase is of weak first order, and between the Higgs- and the Coulomb phase we found a continuous transition.

For two different combinations of the parameters $M^2$ and $\beta$, one in the Higgs phase, one in the confining phase 
(in Fig.~\ref{phasediagram} the two corresponding points are marked with red crosses)   
we now study how observables depend on the chemical potential $\mu$. In Fig.~\ref{obs_vs_mu} we show $\langle U \rangle$,
$\langle |\phi|^2 \rangle$ and $n$, as well as some of the dual occupation numbers versus $\mu$. The top row of plots is
for the point in the Higgs phase ($\beta = 0.75, M^2 = 5.73$) and the bottom row for the confined phase ($\beta = 0.75, M^2 = 4.92$).
 
The general behavior in the Higgs phase (top row in Fig.~\ref{obs_vs_mu}) is the same for all observables: 
They are independent of the chemical potential $\mu$ up to $\mu = \mu_c \sim 2.65$. At this value there is a prominent first order jump 
followed by an essentially linear behavior for $\mu > \mu_c$. This finding is an example of the Silver Blaze behavior, i.e., the
phenomenon that observables are independent of $\mu$ as long as the chemical potential is smaller than the lowest mass
in the spectrum of excitations.  In \cite{silverblaze} it was shown for the charged $\phi^4$ field
that indeed $\mu_c$  very precisely coincides with the mass of the lowest excitation as determined in a spectroscopy calculation. 
The situation is different in the Coulomb phase (bottom row of plots in Fig.~\ref{obs_vs_mu}): There a non-trivial $\mu$-dependence 
sets in already at $\mu = 0$. There is no mass gap, and thus no Silver Blaze type of behavior. 

\section{Spectroscopy with dual variables}

Having discussed results for the bulk observables that can easily be expressed as moments of weighted 
sums of dual variables, let us now come to the question of how to implement $n$-point functions in a dual representation. 
The most general way is to add source terms $\sum_x \phi_x j_x$  to the action (\ref{action_higgs}) of the matter fields. 
Again the dual transformation can be carried out in essentially the same way, with the main difference that one has additional 
dual variables that serve as sources and sinks for matter flux. Thus instead of closed loops, one has open strings as the
dual degrees of freedom for the matter fields. 
The situation is somewhat simpler if one considers $n$-point functions. As an example we discuss the dual representation 
of 2-point functions for the charged $\phi^4$ field, i.e., the theory described by the action $S_M$ of Eq.~(\ref{action_higgs}) 
with $U_{x,\nu} = 1$.

In the conventional representation the 2-point functions are given by
\begin{equation}
\langle \phi_y \, \phi_z^* \rangle \; = \; \frac{1}{Z} \int D[\phi] \, e^{-S_M} \, \phi_y \, \phi_z^* 
\; \equiv \; \frac{1}{Z} \, Z_{y,z} \; ,
\end{equation}
where $Z_{y,z}$ is used to denote the partition sum with two field insertions $\phi_y$ and  $\phi_z^*$. The derivation of
Section 2 is easily generalized to find the dual representation of $Z_{y,z}$,
\begin{eqnarray}
Z_{y,z} & = &\! \sum_{\{ \overline{l}, l\}} \; e^{\,- \mu \sum_x l_{x,4} }   \prod_x \delta\!\left(\!\sum_\nu \big[ {l}_{x,\nu}  -  
{l}_{x-\widehat{\nu},\nu}  \big] - \delta_{x,y} + \delta_{x,z} \! \right) 
\label{Zxy}\\
&& \times \;
\prod_{x,\nu}\! \frac{1}{(|l_{x,\nu}| + \overline{l}_{x,\nu})! \, \overline{l}_{x,\nu}!}  \;
P \! \left( \!\! \sum_\nu\!\big[ |l_{x,\nu}| \! + \!  |l_{x-\widehat{\nu},\nu}| + 2( \overline{l}_{x,\nu} +
\overline{l}_{x-\widehat{\nu},\nu}) \big]  + \delta_{x,y} + \delta_{x,z} \right) \! . 
\nonumber 
\end{eqnarray}
At the positions $y$ and $z$ where the source fields $\phi_y$ and $\phi_z^*$ are 
located the index of the weight factors $P(n)$ is increased (see the Kronecker deltas 
in the argument of the $P$), and also the constraints for the 
$l$-fluxes have additional terms at the sites $y$ and $z$, such that a flux can start at $y$ and end at $z$.
Thus in the dual representation for $Z_{y,z}$ the set of configurations of allowed $l$-flux 
consists of closed loops and a single open line of flux connecting the sites $y$ and $z$.

For an efficient evaluation of  $Z_{y,z}$ it was proposed \cite{wolff2point} to consider 
a generalized partition function ${\cal Z}$  where one sums over all possible positions of the field insertions,
${\cal Z} \; \equiv \; \sum_{u,v} Z_{u,v}$. The configurations that constitute ${\cal Z}$ consist of closed loops of flux 
plus a single flux line with open ends at arbitrary sites $u$ and $v$ of the lattice. For the Monte Carlo simulation of 
${\cal Z}$ we may re-use the standard worm algorithm, and view every step of the worm as an admissible configuration 
contribution to ${\cal Z}$ (not only the configurations where the worm has closed that contribute to $Z$). 
The two point functions are then obtained as
\begin{equation}
\langle \phi_y \, \phi_z^* \rangle \; = \; \frac{Z_{y,z}}{Z} \; = \;
\frac{ \langle \, \delta_{u,y} \, \delta_{v,z} \, \rangle_{\cal Z} }{ \langle \, \delta_{u,v} \;
P(f_u)/P(f_u + 2) \, \rangle_{\cal Z}} \; ,
\label{2pointflux}
\end{equation}
where $\langle .. \rangle_{\cal Z}$ denotes the expectation value with respect to the enlarged ensemble ${\cal Z}$. Here 
$f_u = \sum_\nu\!\big[ |l_{u,\nu}| \! + \!  |l_{u-\widehat{\nu},\nu}| + 2( \overline{l}_{u,\nu} + \overline{l}_{u-\widehat{\nu},\nu}) \big] $ 
denotes the combined $l$- and $\overline{l}$-flux at site $u$ that enters the weights $P$, and the 
reweighting with $P(f_u)/P(f_u + 2)$ in (\ref{2pointflux}) 
is necessary to correctly obtain the original partition sum $Z$ in the denominator of (\ref{2pointflux}). 

\begin{figure}[t]
\centering
\includegraphics[width=100mm,clip]{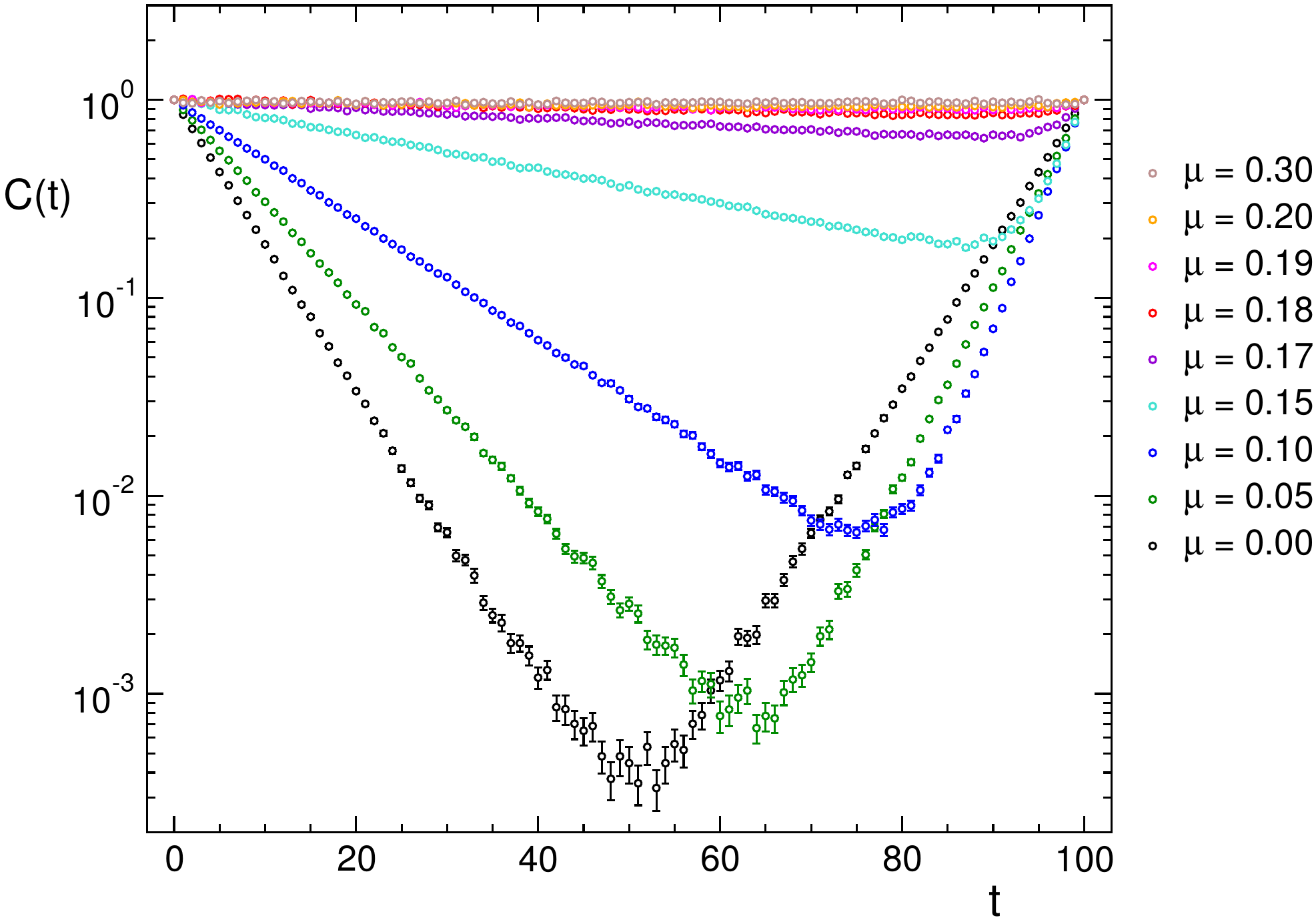}
\caption{Zero momentum propagator as a function of euclidean time $t$ for the 
charged $\phi^4$ field at zero and non-zero chemical potential.}
\label{propagator}
\end{figure}

Based on these techniques, in \cite{dualspectroscopy1,dualspectroscopy2} 
the 2-point functions  for the charged $\phi^4$ field were
evaluated for finite $\mu$. In Fig.~\ref{propagator} we show the results for the 2-point function projected 
to zero momentum as a function of the euclidean time separation $t$. The results for different values of 
$\mu$ are compared.  
While for $\mu = 0$ the propagator has the conventional form of symmetrical forward and backward propagation 
in time, this is different for non-zero chemical
potential. As is obvious from Fig.~\ref{propagator}, for $\mu > 0$ one finds a different slope for forward 
and backward propagation. The asymmetry increases with increasing $\mu$ until $\mu$ becomes so large
that Bose-Einstein condensation sets in (here at $\mu \sim 0.17$), and the propagator becomes flat. The propagators 
were analyzed quantitatively and it was found that the exponents that describe the decay for forward
and backward propagation are given by $m \pm \mu$, where $m$ is the renormalized mass extracted
from the $\mu = 0$ propagator. The results from this simple example show that there is no principal 
obstacle for dual spectroscopy at finite density -- at least for the simple theory studied here.

\section{Dualized models as reference systems}

As we have already outlined in the introduction an important application of finite density systems where a simulation
with dual methods is possible is their use as reference systems for testing and assessing other approaches 
to finite density such as series expansions, stochastic methods et cetera. In this section we report 
about two such comparisons.

An interesting new development for stochastic techniques is the formulation and simulation of
quantum field theories on a Lefschetz thimble \cite{lefschetz1}--\cite{lefschetz5}, a manifold that generalizes 
the notion of curves of steepest descent to higher dimensions. As for the complex Langevin approach the 
degrees of freedom are generalized to complex numbers and the Lefschetz thimble is then defined as the path 
in the complex plane where the imaginary part of the action is constant and the corresponding constant phase 
of the Boltzmann factor cancels. 

The development of the method has reached a stage where simulations of various lattice models were presented, 
one of them being the charged $\phi^4$ model at finite chemical potential \cite{lefschetz3}, i.e., the theory described 
by the action $S_M$ of Eq.~(\ref{action_higgs}) with $U_{x,\nu} = 1$. For this system one can compare to 
the results from the dual simulation \cite{silverblaze} and in this way assess the newly developed method. 

\begin{figure}[t]
\centering
\includegraphics[width=105mm,clip]{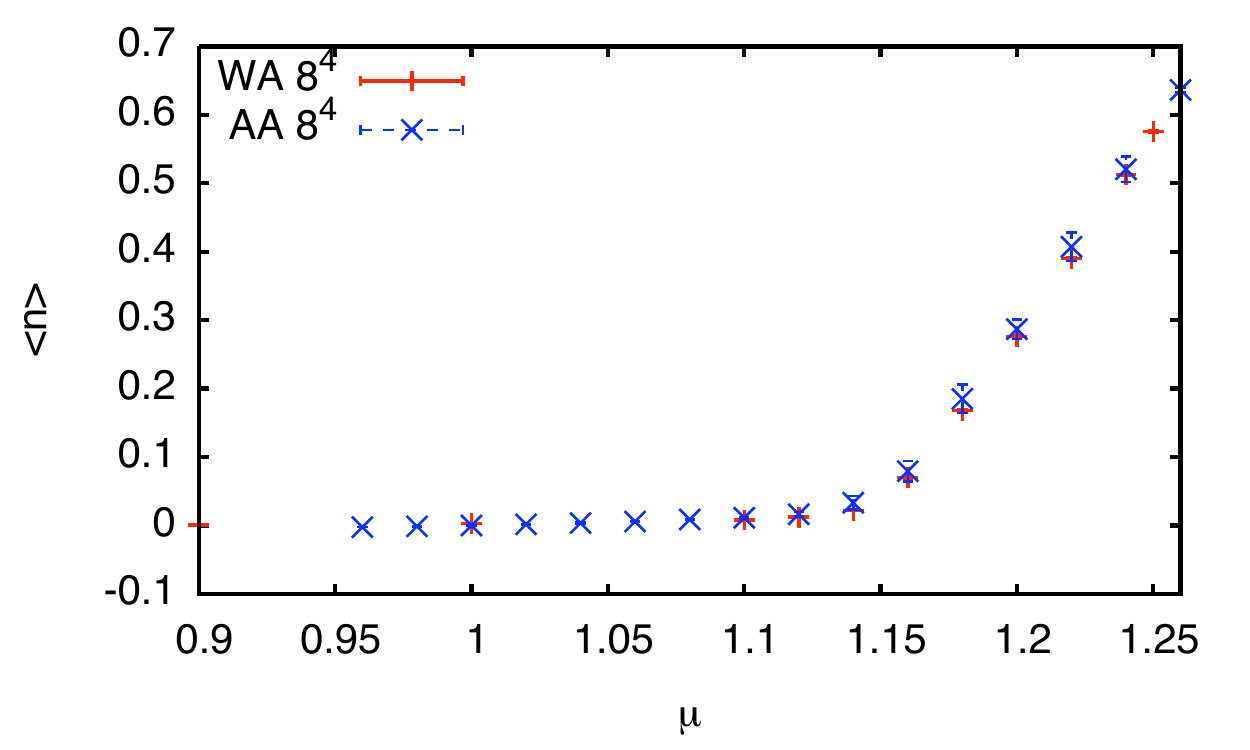}
\caption{Comparison of results from the Lefschetz thimble approach (labelled AA) to the outcome of a dual simulation (WA). The plot from 
\cite{lefschetz3} shows the particle number as function of $\mu$ 
for the charged $\phi^4$-field.}
\label{lefschetzcompare}
\end{figure}

In Fig.~\ref{lefschetzcompare} we compare the results for the particle number density $\langle n \rangle$ as a function of $\mu$. 
The curve shows the onset of condensation at $\mu \sim 1.14$ (the transition is somewhat rounded since the
simulation is for a $8^4$ lattice where finite temperature effects are non-negligible). It is obvious from the plot, that the results
from the Lefschetz thimble approach and from the dual simulation agree quite well. Another example of a comparison of 
stochastic methods to a dual simulation is the test of complex Langevin techniques in the SU(3) spin model 
\cite{langevin8,su3_1,su3_2,su3_3}.

A further interesting application of systems accessible to dual techniques is their use as reference models for various expansions
in the chemical potential $\mu$. The idea is to expand around the $\mu = 0$ theory, where the complex action problem 
is absent and thus the expansion coefficients can be computed with conventional simulations. The simplest such expansion is the
Taylor series, i.e., the partition function $Z$ (or also the free energy) is expanded in $\mu^2$ ($Z$ is an even function in $\mu$). 
However, it is not a-priori clear whether this choice of functions to expand in is optimal. An alternative \cite{fugacity1,fugacity2}  
from statistical mechanics is the fugacity expansion where $Z$ is expanded in a Laurent series 
in the fugacity variable $z = e^{\, \mu \beta}$. 

To assess the two expansions, fugacity and Taylor series, 
they were evaluated in the $\mathbbm{Z}_3$ spin model, where a
dual simulation is available for reference \cite{z3spin1}--\cite{z3spin4}.  The system is a simple effective
theory for the Polyakov loop with an action given by
\begin{equation}
S  \;  = \; - \sum_x \left[ \tau  \sum_{\nu = 1}^{3} \left[ P_x P_{x + \hat{\nu}}^*  + c.c. \right] + \kappa \, e^\mu \, P_x + 
\kappa \, e^{\, -\mu}P_x^* \right] ,
\label{Z3effectiveaction}
\end{equation}
where the sum runs over the sites $x$ of a 3-d  lattice and the Polyakov loop variables
$P_x$ are elements of $\mathbb{Z}_3 = \left\lbrace 1, e^{\pm 2 i \pi /3} \right\rbrace$. $\tau$ is an increasing function of the temperature,
$\kappa$ a function that decreases with increasing quark mass and $\mu$ is the chemical potential. For $\mu \neq 0$ 
the action is complex.

\begin{figure}[t]
\centering
\includegraphics[width=75mm,clip]{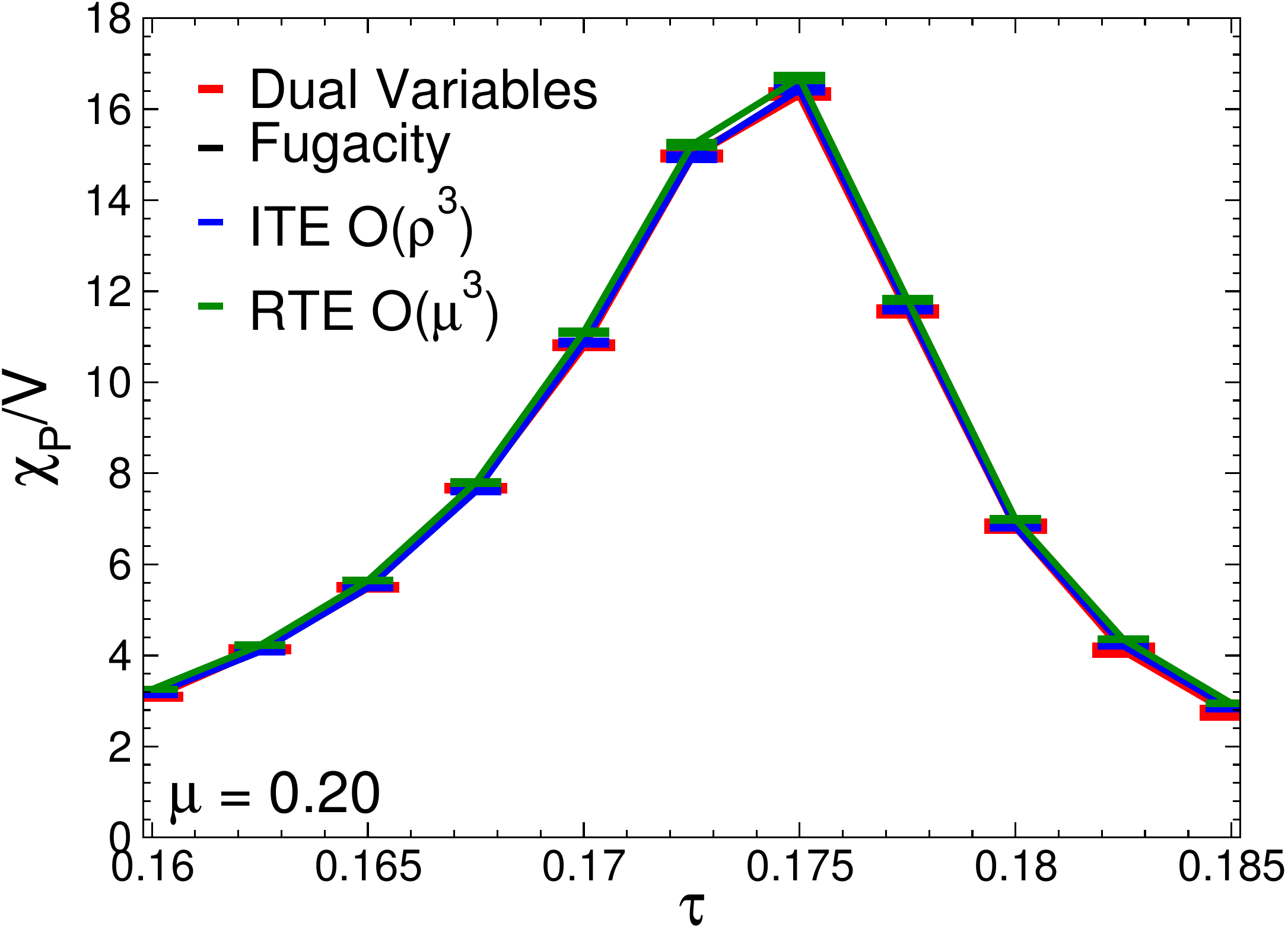}
\includegraphics[width=75mm,clip]{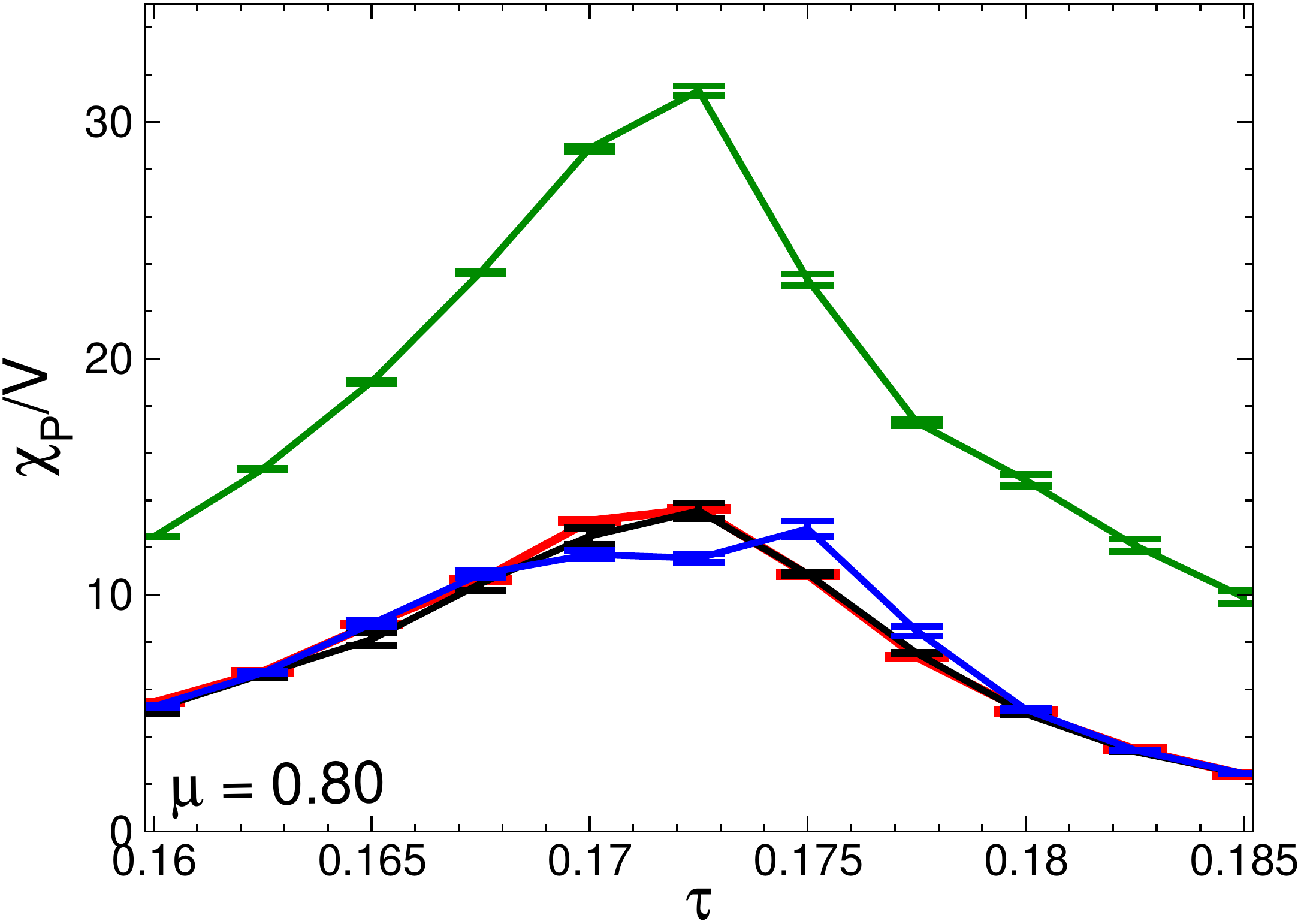}
\caption{Comparison of various series expansions in $\mu$ for susceptibility 
$\chi_P/V$ in the $\mathbbm{Z}_3$ spin model to the results from a dual simulation. 
We show $\chi_P$ from fugacity expansion, improved Taylor expansion (ITE) and 
regular Taylor expansion (RTE) as a function of the temperature parameter
$\tau$ for two different values of the chemical potential, $\mu = 0.2$ (lhs.\ plot) and $\mu = 0.8$ (rhs.).}
\label{seriescompare}
\end{figure}

The Taylor expansion, the fugacity expansion and a modified Taylor expansion (power series in $e^{\pm \mu} - 1$) are worked out 
easily and can be compared to the results of the dual approach. Fig.~\ref{seriescompare} shows the results for the Polyakov loop
susceptibility as a function of the temperature parameter $\tau$ at $\kappa = 0.01$ for two different values of
the chemical potential, $\mu = 0.2$ in the lhs.\ plot and $\mu = 0.8$  on the rhs. The figures show that at $\mu = 0.2$ all three series expansions
(Taylor and improved Taylor expansion are up to third order) agree quite well with the dual simulation results, while for $\mu = 0.8$
the regular Taylor expansion is far off the dual curve, the improved Taylor series starts to have problems, and only the fugacity
expansion still reasonably matches the dual results. 

In another project the recent developments for the density of states approach \cite{dos1,dos2,dos3} will be implemented in the 
system (\ref{Z3effectiveaction}) and cross-checked with the dual results. 
We expect that model studies in systems where a dual simulation is possible can help to optimize 
new ideas and strategies for finite density before applying them in considerably more expensive realistic theories. 

\section{Fermions}

In this section we briefly discuss   aspects of dual representations for systems with fermionic matter. As a toy example we 
consider a system of staggered fermions with a simple quartic interaction. The action is given by 
\begin{equation}
S \; = \; \sum_x \left( m \overline{\psi}_x \psi_x +  \sum_\nu \frac{\eta_\nu(x)}{2}
\left[ e^{\, \mu \delta_{\nu 4}} \; \overline{\psi}_x \psi_{x+\hat{\nu}} - e^{\, - \mu \delta_{\nu 4}} \; \overline{\psi}_x \psi_{x-\hat{\nu}} \right] 
- g \sum_{\langle xy \rangle} \overline{\psi}_x \psi_x \,  \overline{\psi}_y \psi_y \right) \; .
\end{equation} 
Here $\overline{\psi}_x, \psi_x$ are single component Grassmann valued fields at the sites $x$ of the lattice, 
$\eta_\nu (x) = (-1)^{\sum_{\alpha = 1}^{\nu -1} x_\alpha}$ is the staggered sign factor which plays the role of the 
Dirac matrices, $m$ the fermion mass, $g$
the quartic coupling and the last sum runs over all neighbors $y$ of the site $x$.

Similar to the case of the bosonic model considered in Section 2, the Boltzmann factor $e^{-S}$ in the path integral 
can be factorized into a product of individual terms that live on links and on sites. Each of these factors can be 
expanded, and since Grassmann variables are nilpotent, the expansion terminates after the second term, such that we get for 
the example of a mass term at site $x$ the series 
$e^{-m \overline{\psi}_x \psi_x} = \sum_{s_x = 0}^1 (-m)^{s_x} \, (\overline{\psi}_x \psi_x)^{s_x}$ and similar for the other terms. 
The expansion indices ($s_x$ in our example) can have only two values, $s_x \in \{0,1\}$,  and thus can be interpreted 
as ''activation indices'' that activate ($s_x = 1$) certain terms or turn them off ($s_x = 0$). For a non-vanishing result
of the Grassmann  integral $Z = \int \prod_x \, d \overline{\psi}_x \, d\psi_x \, e^{-S}$ each $\psi_x$ and each $\overline{\psi}_x$
must be activated exactly once. This leads to constraints for the allowed configurations of the activation indices, which again give
rise to loops of dual variables living on links from the hopping terms, but also insertions of ''monomers'' from the
mass and quartic coupling terms which completely saturate the Grassmann integrals at all sites that they are activated on. 

A serious complication comes from the fact that for the evaluation of the Grassmann integral the individual variables
$\overline{\psi}_x$ and $\psi_x$ have to be brought into a canonical order, which due to the anti-commutating nature
of the Grassmann variables gives rise to additional minus signs. For example every loop of dual flux comes with a
minus sign. Further minus signs that depend on the shape of the loops come from the staggered factors $\eta_\nu(x)$.
Thus although we do have a dual representation of the system (where again the chemical potential enters in a harmless way),
the complex action problem is not solved, since the configurations of the dual variables come with both signs. 
Although examples are known (in particular in low dimensions) where this simple dualization 
leads to a form useful for simulations \cite{my2dferms_1}--\cite{wolff2d_2}, the general perspective is 
not very good for this approach.

An interesting concept is the so-called fermion bag approach \cite{fbagfirst}--\cite{fbaglast}: 
There only the Boltzmann factor of the interaction term
is expanded. If the interaction is such that the corresponding activated terms completely saturate the Grassmann integral
locally (this is the case for our example), 
the sites where the interaction is activated can be removed from the Grassmann integral. What remains is a free 
theory (only hopping and mass terms) on a lattice where sites are removed. The partition 
function turns into an unconstrained sum over all configurations of the activation indices for the fermion interaction,
and the method becomes efficient in a parameter range where many interaction terms are activated and the terms
of the free theory need to be evaluated only on small patches of the lattice (fermion bags). Several
interesting models could be studied with the fermion bag approach and we refer the reader to the lecture series 
\cite{fermionlast} by Chandrasekharan for an overview and the discussion of further aspects of the fermion bag method and related ideas.  

\begin{figure}[t]
\centering
\includegraphics[width=105mm,clip]{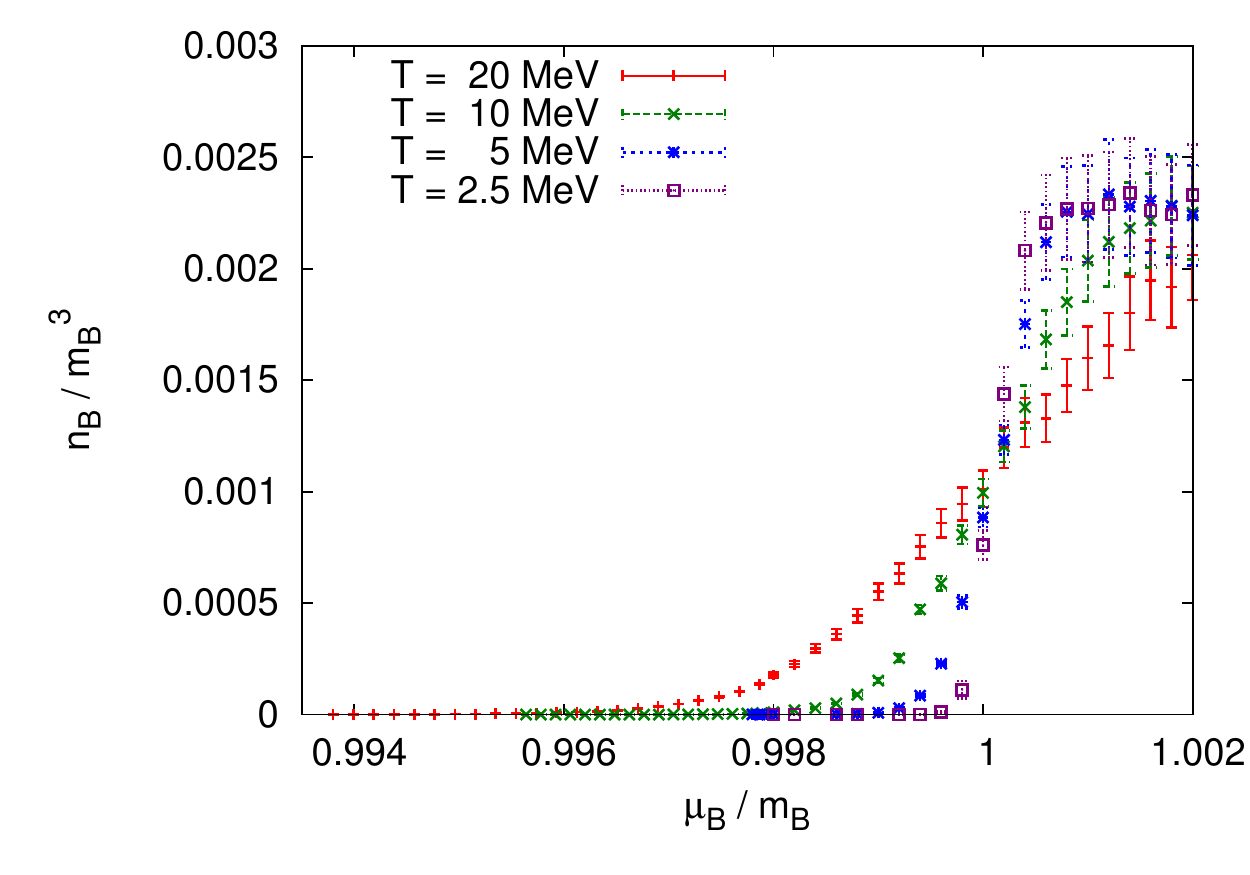}
\caption{Particle number density as a function of the chemical potential for lattice QCD in the strong-coupling / large-mass limit.
(Figure from \cite{owe2}.)}
\label{heavystrong}
\end{figure}

\section{Strong coupling and large mass}

In the first few sections of this contribution we have shown that  for bosonic matter with abelian symmetries the problem of 
solving the complex action problem by mapping the system to dual variables is essentially solved. In the previous section
it was discussed that for fermions the challenge is much harder due to the extra minus signs of the matter loops from the
anti-commutation relations of the fermions and signs from the Dirac matrices. Similarly difficult is the question for a suitable
dual representation of non-abelian gauge fields (for some attempts see \cite{nonabfirst}--\cite{nonablast}). 

It is, however, well known that in the limits of large quark mass $m$ and strong coupling one can compute the leading terms
of a dual representation, and that these leading terms are better behaved from a Monte Carlo point of view: Large mass $m$
suppresses long matter loops, as each link of matter flux costs a weight factor $\propto 1/m$. In a similar way each occupied 
plaquette of a gauge surface costs a factor $\beta$, and the inverse gauge coupling $\beta$ is small in the strong coupling limit.  
Several attempts to study such partially dualized theories in the large-mass/strong-coupling limits can be found in the literature
(see, e.g., \cite{fermionfirst}--\cite{fromm}, \cite{fbag2}, \cite{owe1}--\cite{owe4}).

A particularly elaborate implementation of a dualization in the large-mass/strong-coupling limit has been presented in 
\cite{owe1}--\cite{owe4}. The leading terms of the dual representation in the large-mass and strong-coupling limits
were taken into account, but the coefficients were determined with a matching procedure to optimally reproduce physics
at $\mu = 0$. The resulting dual theory was then studied at finite $\mu$, and as an example of what may be achieved,
Fig.~\ref{heavystrong}  shows the baryon number density as a function of the chemical potential for different temperatures.
One nicely sees the condensation transition becoming sharper as the temperature is decreased.

It is clear that dual simulations in the large-mass/strong-coupling region have considerable limitations -- nevertheless
the results that were obtained recently nicely illustrate physical mechanism that play an important role also
at physical parameter values. 

\section{Concluding remarks}

In this contribution we have discussed recent developments for attempts to overcome the complex action problem of lattice 
field theories by the use of transformations to new variables (''dual variables'') where the partition sum has only real and 
positive contributions. Our main example was the U(1) gauge Higgs model which is characteristic for all 
abelian gauge Higgs models, a class of systems where the finite density complex action problem is essentially solved. 
Dual representations have the form of conserved discretized fluxes for matter fields and surfaces that are either closed or
bounded by matter flux for the gauge degrees of freedom. We discussed new algorithmic ideas for dual simulations
and typical examples for results that were obtained, including results for 2-point functions and spectroscopy. 

For fermion systems and non-abelian symmetries the progress is not as far advanced, although the general
structure of loops and surfaces is the same. For fermions the Grassmann variables and the Dirac matrices
give rise to additional minus signs for the matter flux and the non-abelian nature of the gauge fields prevents
one from the reordering of the link variables that led to a successful dualization in the abelian case. 
We stress, however, that the abelian and bosonic examples show that dual representations are not unique and 
thus there is hope that further interesting ideas and developments might also be possible for fermions and 
systems with non-abelian symmetries.

\end{document}